\documentclass[twocolumn,superscriptaddress,nofootinbib,preprintnumbers,showpacs,tightenlines,notitlepage]{revtex4}
\usepackage[latin1]{inputenc}
\usepackage{graphicx}
\usepackage{amssymb}
\usepackage{color}
\usepackage{float}
\usepackage{amsmath}
\usepackage{amsfonts}
\usepackage{dcolumn}
\usepackage{hyperref}
\usepackage{amsthm}
\theoremstyle{definition}
\newtheorem{definition}{Definition}[section]
\usepackage{color}

\def\uudot{\dot{u}}
\def\3nab{\tilde{\nabla}}

\def\la {\langle}
\def\ra {\rangle}

\def\be {\begin{equation}}
\def\ee {\end{equation}}
\def\ba {\begin{eqnarray}}
\def\ea {\end{eqnarray}}

\newcommand{\bra}[1]{\left(#1\right)}

\newcommand{\sfr}[2]{{\textstyle\frac{#1}{#2}}}

\newcommand{\lc}{\varepsilon}
\newcommand{\lb}{\{}
\newcommand{\rb}{\}}
\newcommand{\E}{{\mathcal E}}

\newcommand{\barray}{\begin{array}}
\newcommand{\earray}{\end{array}}
\newcommand{\e}{e}
\newcommand{\N}{N}

\newcommand{\del}{\nabla}

 \newcommand{\nab}{\nabla}

\newcommand \ep {\epsilon}

\newcommand \om {\omega}
\newcommand{\na}{\nabla}
\newcommand{\udot}{{\mathcal A}}
\begin{document}

\title{Universality and Criticality in Mass-less Scalar Field Collapse}

\author{Koushiki}
\email{koushiki.malda@gmail.com}
\affiliation{International Centre for Space and Cosmology, School of Arts and Sciences, Ahmedabad University, Ahmedabad, GUJ 380009, India}

\author{Rituparno Goswami}
\email{Goswami@ukzn.ac.za}
\affiliation{Astrophysics and Cosmology Research Unit, School of Mathematics, Statistics and Computer Science, University of KwaZulu-Natal, Private Bag X54001, Durban 4000, South Africa.}

\author{Pankaj S. Joshi}
\email{psjcosmos@gmail.com}
\affiliation{International Centre for Space and Cosmology, School of Arts and Sciences, Ahmedabad University, Ahmedabad, GUJ 380009, India}

\begin{abstract}
In this paper, we observe the collapse of a mass-less scalar field covariantly. We show that the strengths of the collapsing and dispersing modes of this scalar field will decide whether the collapse will end up in a black-hole or disperse. We find a locally naked null singularity as a critical case between these two and confirm that there is a single dimensionless parameter that determine the end state. This work is ansatz-independent, hence, true for all mass-less scalar field families. We also show that the geometrical mass of these singularities go to zero.

\end{abstract}


\maketitle
\section{Introduction}
In 1969, Roger Penrose gave his famous {\it Cosmic Censorship Conjecture (CCC)}\cite{CCC}, which restricts the causal connection from the neighbourhood of a singularity in all future directions. Since its introduction, it has remained a focal point of debate amongst gravitational physicists. On one hand, CCC gave birth to a whole new branch of physics: black-hole physics, in which the validity of CCC in indispensable. On the other, quests to find collapse solutions ending up in black-holes and proving CCC ended up finding out crucial counter-examples. In these, families of intextendible causal geodesics exist in the neighbourhood of the singularity which are past incomplete and future complete, while {\it Strong} CCC prohibits against the existence of such geodesics and {\it Weak} CCC prohibits these geodesics from being future complete. These objects were later named as {\it naked singularities}, {\it fire-balls} or {\it quantum-stars}. The existence of these objects opened up a Pandora's box full of new problems in front of the physics community by the removal of global-hyperbolicity from the space-time manifolds, or that these structures fall outside the manifold structure themselves. The existence of such astrophysical objects also opened up the possibility of observations similar to that of the initial cosmological singularity.  Despite all these attempts in both sides, a concrete mathematical proof or disproof of CCC still eludes us. \\\\
While CCC does not allow the existence of naked singularities, naked singularities do not contradict the existence of black-holes. Rather, pivotal works that postulate naked singularities as a non-trivial subset of solutions of {\it Unhindered Gravitational Collapse} (UGC), give the other subset of solutions consisting of black-holes as the end-state of UGC. Amongst these works, are those where the appearance of {\it apparent horizon} (AH) was delayed to produce naked singularities by  the introduction of inhomogeneity in the initial density configurations of pressureless dust\cite{Joshi:1991aa, Joshi:1993zg, Joshi:1993jg, Joshi:1994st, Joshi:1997kx, Joshi:2004tb, Goswami:2004gy, RG, Goswami:2002he}. The introduction of pressure in the initial configuration also results in the delaying or non-occurrence of AH in the final-state as well \cite{RG, JMN11, Joshi:2013dva}. While some might argue about the physical relevance of the matter-fields in these works, but the initial configurations described in  these works are type-I matter-fields and follow energy-conditions \cite{HE}. Also, these matter fields are regularly used to explain multiple physical phenomena in cosmological and astrophysical studies, starting from structure-formation to star formation and collapse.\\\\
On the topic of physical relevance of the matter fields, there is no better candidate than a real scalar field $\Psi: \mathcal{M} \to \mathbb{R}$, because it is constituted from a fundamental  Lagrangian and abides by the Klein-Gordon equation. Despite its simplicity of formulation, it can elegantly represent a large number of physical matter fields, including but not limited to, null fluids and perfect fluids (with an additional structure of a potential $V(\phi)$). The validity of CCC has been addressed by following the collapse of scalar fields as well, in general for spherically symmetric space-times. Christodoulou has shown that in the set of initial data, there exists a positive measured subset which forms a naked singularity from a mass-less scalar field (MSF) at the end-state of an UGC \cite{Christ, Christ1}. His claim has been verified for massive scalar fields as well \cite{Goswami:2007}. Further, for a homogeneous and isotropic space-time, these works have been generalised and a class of naked singularities has been shown to be {\it strong} \cite{Tipler}, as well as visible from an assymptotic observer \cite{KK1}. \\
Another intriguing aspect of MSF collapse was observed by Choptuik in 1993\cite{Choptuik}. In this numerical study he found out for a one-parameter$(p)$ family of initial configurations, the scalar field produces a critical mass black-hole. In the phase-space of this parameter, the end-state solution is a black-hole if only the parameter value is close to a critical value $(p^*)$ and away from this value, the scalar field disperses without producing a singularity. He also observed that the mass of the black-hole is universal in all these initial configurations and it has a power-law relationship with this parameter $M_{BH} \propto {(p-p^*)}^\gamma$ . He found the value of $\gamma$ to be universal as well. Further numerical studies \cite{Abrahams, Hamad:2004sw, Garfinkle:1995zj} also confirm this result. To be noted, all these studies confirm a scale-invariance up to fractal-sizes, realisable in numerical computation as well. Hod and Piran \cite{Hod:1996ar} observed a fine-structure behaviour in the exponent $\gamma$. Grafinkle and Duncan \cite{Garfinkle1998} observed the relationship between the maximum curvature at the central singularity and the distance of the parameter from its critical value, which they found to be analogous to Choptuik's work. \textcolor{black}{Brady \cite{Brady1995} showed that this collapse is self-similar.} While all these works are numerical, Gundlach after giving a review in \cite{Gundlach:2002sx}, gives a probable analytical approach to observe criticality in a MSF collapse. This discussion is physically relevant as black-holes have recently gained candidature as dark-matter particles. One of the postulated formation procedures of primordial black-holes is critical collapse of MSFs \cite{Carr2021}.\\\\
In the present work, we give an analytical formulation of observing the critical collapse of a MSF. We perform this without choosing any particular co-ordinate basis, rather we just exploit the symmetries of the space-time to perform the calculation covariantly. We find out that this is the only way to proceed and elaborate the reasons subsequently. We show that for a collapsing scalar field, the strengths of the collapsing and dispersing modes are the only deciding factors which predict the end-state: a black-hole or a dispersal. At the boundary of these two cases, we also find a locally naked singularity. We confirm this by calculating the slope of the apparent horizon near the central singularity, as prescribed in \cite{Joshi:1993zg, Goswami:2006ph, Hamid:2017, Koushiki:2025uqr}. Our approach, being ansatz independent, renders the process of handpicking families of MSFs redundant. We also confirm that the geometrical mass of the singularities go to zero, as observed previously. Furthermore, we confirm the scale-independence of the single parameter, which determines the slope of the apparent horizon in the neighbourhood of the central singularity. The structure of this paper is as follows:

In Sec.(\ref{tetrad}), we introduce the semi-tetrad formalism as a basis of our work. In Sec.(\ref{1+3}), we give the local covariant decomposition of time-orientable space-times. In Sec.(\ref{1+2}), we show, how under another spatial-symmetry on top of time-orientability, the space-time can be further decomposed locally. We obtain, here, the irreducible set of thermodynamic scalars to describe these systems completely if the direction of spatial symmetry is along the radial direction. The broader class of space-times are called LRS-II, of which spherical symmetry constitutes a sub-class. We perform the rest of our calculations in this only.  In sec.(\ref{MSFC}), we describe the dynamical behaviour of a MSF in a double-null basis.
Then with the help of covariant decomposition, introduced in the previous section, we derive the propagation and the evolution equations using doubly contracted Bianchi identities and the Ricci identity. With these, we derive the equations for the apparent horizon and the necessary conditions to achieve the same. We also specify our reason for choosing to work co-ordinate independently here. In Sec.(\ref{MisnerSharpAH}), we calculate the slope of the apparent horizon at the central singularity. With this, we arrive at three different cases, dictated by the strength of the collapsing and dispersing modes of the scalar field. Namely, dispersal ending in a regular end-state, a black hole and a locally naked singularity. We show here that this locally naked singularity is null in nature. We show that these phenomena are scale-independent, in Choptuik's own words {\it echoing}, and the geometrical masses of the obtained singularities are {\it zero}.\\\\

We will use natural units with $c=8\pi G=1$ throughout this paper. The different derivatives will be expressed as: $\nabla \equiv$ co-variant derivative, $\dot{x} \equiv$ directional derivative of $x$ in time-like direction, $\hat{x} \equiv$ directional derivative of $x$ along radial direction, $\delta x \equiv$  directional derivative of $x$ along the orthogonal direction of these two. The signature scheme is $(-,+,+,+)$. The conventional notations to define symmetry and anti-symmetry are respectively:  
\be
T_{(a b)}= \frac{1}{2}\left(T_{a b}+T_{b a}\right)\;,\qquad T_{[a b]}= \frac{1}{2}\left(T_{a b}-T_{b a}\right)\,.
\ee


\section{Covariant Formalisms in Semi-Tetrad Systems}\label{tetrad}

Time-orientable space-times can be locally decomposed into $3$-space, foliated on $\mathbb{R}$, denoting time. This can be done using the tetrad decomposition by Newman and Penrose (NP), the $3+1$ decomposition developed by Arnowitt, Deser and Misner (ADM) or $1+3$ covariant approach by Ehlers and Ellis. If the space-time possesses an additional symmetry along any of the spatial directions, the space can be further decomposed into sheets along that ''special'' spatial direction and we get $1+1+2$ covariant decomposition, which is local as well. Amongst these, the ADM decomposition is not fully covariant. We give a brief description of these formalisms in the next two subsections.

\subsection{Local 1+3 covariant decomposition}\label{1+3}

The tangent to the worldline of a time-like observer is the four-velocity $u^a$ of the observer and is orthogonal to a space-like hypersurface. So, the space-time manifold can be locally decomposed into $1$ time-like congruence, orthogonal to the $3$-space \cite{EllisCovariant}: $\mathbb{R} \otimes \mathcal{V}$, where, $\mathcal{V}$ is the space-like hypersurface. The local nature of this decomposition, takes away the necessity of the existence of a Cauchy surface to study the time-like congruence. Hence, it becomes exceptionally useful to study singular space-times. So, any vector $X^a$ can be projected along the space-like hypersurface by contracting it with: 
\be h^b_{\;\;a} = u^b u_a +g^b_{\;\;a}.\ee
With this, the  \textit{covariant time derivative} of a tensor $M^{a..b}{}_{c..d}$ is: \be
\dot{M}^{c..d}{}_{a..b} = u^{e} \nab_{e} {M}^{c..d}{}_{a..b}\;,
\ee
along the observers' worldlines.
Similarly, the \textit{covariant derivative along space-like directions}, $D$, can be achieved by contracting it with the orthogonally projected tensor $h^a_{\;\;b}$: 
\be\label{orth3+1} D_{l}{M}^{c..d}{}_{a..b} = h^f{}_a
h^c{}_p...h^g{}_b h^d{}_q h^r{}_l \nab_{r} {M}^{p..q}{}_{f..g}\;,
\ee 
and it is projected on all the dummy indices. The volume element on the 3-surface is:
\ba
\label{eps1}
\ep_{a b c}&=&-\sqrt{|g|}\delta^0_{[a} \delta^1_b \delta^2_c \delta^3_{d]} u^d\; \text{and} \\
\label{eps2}
\ep^{a b c} \ep_{d e f} &=& 3! h^a_{[d} h^b_e h^c_{f]}\;, \ep^{a b k} \ep_{d e k} = 2! h^a_{[d} h^b_{e]}\;.
\ea
It is needless to say that the volume element is a 3-form and Eq.(\ref{eps2}) is a natural consequence of its definition. Orthogonal projection of a vector is written as:
\be
W^{\la b \ra} = h^{b}{}_{a}W^{a}~,
\ee
and the orthogonally
\textit{projected symmetric trace-free} PSTF part of tensors (expressed by curly brackets): 
\be
 M^{\lb cd \rb} = \left[
h^{(c}{}_a {} h^{d)}{}_b - \frac{1}{3} h^{cd}h_{ab}\right] M^{ab}\;.
\label{PSTF} 
\ee 
Orthogonal here means normal to the time-like congruence. With this, the covariant derivative of $u^a$ can be decomposed into its irreducible scalar-vector-tensor parts as: 
\be
\nabla_au_b=-\mathcal{A}_au_b+\frac13h_{ab}\Theta+\sigma_{ab}+\ep_{a b
c}\om^c\;, 
\ee 
where, acceleration $\mathcal{A}_a=\dot{u}_a$,
expansion $\Theta=D_au^a$, shear $\sigma_{ab}=D_{\la a}u_{b \ra}$ and vorticity $w^a=\ep^{a b c}D_bu_c$. In a similar manner, Weyl tensor can also be decomposed:
\be 
E_{ab}=C_{abcd}u^cu^d=E_{\la ab\ra}\;;\;
H_{ab}=\frac12\ep_{acd}C^{cd}{}_{be}u^e=H_{\la ab\ra}\;, 
\ee 
where, $E_{ab}$ and $H_{ab}$ are the electric and magnetic parts of it. Both of these are completely anti-symmetric. Just like these, the stress-energy tensor can be decomposed into its irreducible parts:
\be\label{STRESS}
T_{ab}=\mu u_au_b+q_au_b+q_bu_a+ph_{ab}+\pi_{ab}\;,
\ee
with $\mu=T_{ab}u^au^b\equiv$ the energy density, $p=(1/3 )h^{ab}T_{ab}\equiv$ isotropic pressure, $q_a=q_{\la a\ra}=-h^{c}{}_aT_{cd}u^d\equiv$  the heat flux and $\pi_{ab}=\pi_{\la ab\ra}\equiv$ anisotropic stress. For detailed calculations of these decompositions, one should refer to \cite{ellis_van_elst}. 

\subsection{Local 1+1+2 covariant decomposition}\label{1+2}

In a space-time, possessing an additional symmetry along any of the three spatial-directions, locally the space-time can be further broken down to $\mathbb{R} \otimes \mathbb{R} \otimes \mathcal{S}$, where $\mathcal{S}$ is the space-like $2$-surface orthogonal to both the time-like and the preferred space-like directions. This is the $1+1+2$ formalism, developed by Clarkson and Barrett. Amongst many other applications, it has been used extensively to study perturbations of black holes \cite{Clarkson:2002jz,Betschart:2004uu,Clarkson:2007yp}. Spherical symmetry is one of the cases where, this decomposition can be applied. Under this scheme, the geometrical equations become algebraic, compared to tensorial in the $1+3$ decomposition.  
 
In such space-times, $\exists$ a space-like $e^a$ orthogonal to $u^a$ and the other two space-like directions. So,  
$\e^a\e_a=1,~u^a\e_a=0$. So, the projection tensor, orthogonal to both $e^a$ and $u^a$ becomes:
\begin{eqnarray} \label{1+1+2}
\N_a^{~b}&=&
h_a^{~b}-\e_a\e^b=g_{a}^{~b}+u_au^b-\e_a\e^b, \nonumber\\ 
\e^a\N_{ab}&=&0=u^a\N_{ab},\;\; N^a{}_a=2.
\end{eqnarray}
The tensor $N_{ab}$ defined on the $2$D orthogonal hypersurface, is called the \textit{sheet}. In spherical symmetry, it is the two-sphere. The volume element on this $2$-surface is the $3$-volume projected along the preferred spatial direction:
\be \lc_{ab}\equiv\ep_{abc}\e^c = u^d\eta_{dabc}e^c, \ee
and evidently $\lc_{ab}\e^b=0=\lc_{(ab)}$, as this is orthogonal to $e^a$.
A vector on the space-like $3$-surface $V^a$ can be split into its irreducible scalar-vector parts: \ba 
v_a&=&V\e_a+V_{a},~~~\mbox{where}~~~V\equiv V^a\e_a\
,\nonumber\\&&~~~\mbox{and}~~~V_{a}\equiv \N_{ab}V^b\equiv v_{\bar a}\label{psia}.
\ea Here, $V$ lies along $e_a$. The bar over the index signifies the projection along $N_{ab}$ and hence, $v_{\bar a}$ lies on the sheet.
A tensor can also be split into its irreducible parts similarly:
 \be
\rho_{ab}=\rho_{\langle
ab\rangle}=P\bra{\e_a\e_b-\sfr12\N_{ab}}+2P_{(a}\e_{b)}+P_{{ab}}\
, \label{tensor-decomp} 
\ee where
 \ba
P&\equiv &\e^a\e^b \rho_{ab}=-\N^{ab}\rho_{ab}\ ,\nonumber\\
P_a&\equiv &\N_a^{~b}\e^c\rho_{bc}=P_{\bar a}\ ,\nonumber\\
P_{ab}&\equiv &
\bra{\N_{(a}^{~~c}\N_{b)}^{~~d}-\sfr{1}{2}\N_{ab}\N^{cd}}\rho_{cd}
\equiv P_{\lb ab\rb}\ . \label{PSTF-TT}
\ea 
 The covariant derivatives along $e^a$  can also be defined now, just like in $1+3$ decomposition: 
\be
\hat{\rho}^{c..d}{}_{a..b} = e^{k}D_{k}{\rho}^{c..d}{}_{a..b}~.
\ee
In this case, the orthogonal surface is the 2-sheet and the orthogonal projection is similar to Eq.(\ref{orth3+1}) with $h_{ab}$s replaced by $N_{ab}$s:
\be
 \delta_{l}{\rho}^{c..d}{}_{a..b} = N^f{}_a
N^c{}_p...N^g{}_b N^d{}_q N^r{}_l \nab_{r} {\rho}^{p..q}{}_{f..g}\;,
\ee 
The projection runs over all the dummy indices. With these at hand, the acceleration on $\mathcal{V}$ can be further broken down into:
\ba
\uudot^a&=&\udot \e^a+\udot^a,
\ea
with $\mathcal{A}$ along $e^a$ and $\mathcal{A}^a$ on the sheet. All other kinematic vector and tensor quantities can similarly be broken down into their irreducible parts:
\ba
\omega^a&=&\Omega \e^a+\Omega^a,\\
\sigma_{ab}&=&\Sigma\bra{\e_a\e_b-\sfr{1}{2}\N_{ab}}+2\Sigma_{(a}\e_{b)}+\Sigma_{ab},
\ea 
and the full covariant derivative of $\e_a$ and $u_a$ is now written as:
\ba\label{del_e}
 \del_a\e_b&=&-\udot u_au_b+\bra{\Sigma+\sfr13\Theta}\e_a u_b+\sfr{1}{2}\phi\N_{ab},\\
\label{del_u}
\del_au_b&=&
-u_a\e_b\udot +\e_a\e_b\bra{\sfr13\Theta+\Sigma}+\frac12\N_{ab}\bra{\sfr23\Theta-\Sigma}, \nonumber \\
\\
\text{with}\;\; \udot &=& \e_a\dot{u}^a \;, \;\; \Sigma = \e_a e_b\sigma^{ab}\;,\nonumber\\
\Theta&=& D_a u^a\;,\; \phi = \delta_a e^a\;.
\ea
The electric and magnetic parts of the Weyl tensor are further decomposed into:
\ba\label{weyl1}
E_{ab}&=&{\cal E}\bra{\e_a\e_b-\sfr{1}{2}\N_{ab}}+2{\cal E}_{(a}\e_{b)}+{\cal E}_{ab}, \\ 
H_{ab}&=&{\cal H}\bra{\e_a\e_b-\sfr{1}{2}\N_{ab}}+2{\cal H}_{(a}\e_{b)}+{\cal
H}_{ab}.
\ea

These equations are written using 
Eq.(\ref{psia}), Eq.(\ref{tensor-decomp}) as followed from \cite{Clarkson:2007yp}. The $1+1+2$ covariant-decomposition is especially suitable for a space-time which has a continuous isometry group defined along a spatial axis. These space-times are called \textit{locally rotationally symmetric} (LRS). LRS spacetimes are isotropic about this spatial axis. \textit{All} 2-vectors and 2-tensors vanish and there are no preferred directions in the sheet.  For spherically symmetric space-times, this isometric group is $SO(3)$. In this case, the magnetic Weyl tensor also vanishes and we are left with the electric Weyl tensor only:
\be 
\label{weyl2}\mathcal{E} = E^{ab}e_a e_b \;.
\ee 
And the fluid variables or the thermodynamic quantities are similarly broken into their irreducible forms using Eq.(\ref{1+1+2}) and only their scalar parts produce {\it non-zero} contributions:
\ba \label{Thermo_sc}
\mu &=& T_{ab} u^a u^b, \;\; p=\frac13 t_{ab}h^{ab},\;\; \mathcal{Q} = \mathcal{Q}_a e^a=-h^c_a T_{cd} e^a u^d,\nonumber\\
\Pi &=& \left[\left(h^c_{(a} h^d_{b)} -\frac13 h_{ab}h^{cd}\right) T_{cd}\right] e^a e^b.
\ea
To summarise, in this spatial symmetry it is very convenient to analyse the kinematic equations and these can be written with nothing but a set of scalars which completely describe the system:
\ba
\mathcal{D} \equiv \bra{\Theta , \mathcal{A}, \Sigma, \mathcal{E}, \mu, \phi, p, \Pi, \mathcal{Q}}\;.
\ea
 Since LRS spacetimes are isotropic about all rotations around $\e_a$, \textit{all} $2$-vectors and $2$-tensors vanish on the sheet perpendicular $\e_a$. So, the covariant description of this system only entails scalars on this $2$-sheet. Detailed calculations can be looked up from \cite{Clarkson:2007yp}. In the LRS class, the rotationless space-times are called LRS class II, which contains the spherically symmetric space-times. We will be considering only this class in the current work.

\section{Mass-less scalar field in a spherically symmetric space-time}\label{MSFC}

Choptuik, in his phenomenal work of $1993$ \cite{Choptuik}, showed that the collapse of families of MSFs, depending on a single parameter value, conformally end up in black-holes. He also showed that the masses of the produced black-holes for \textit{all} these families have a universal behaviour. The results of this work has been reinforced later in various numerical studies \cite{Abrahams, Hamad:2004sw, Garfinkle:1995zj, Hod:1996ar, Garfinkle1998, Gundlach:2002sx, Brady1995}. We show analytically in the next sections that collapse of a MSF, indeed, universally ends up in a black-hole for a critical parameter value. We also discuss the value of the same parameter for which the end-state of such a collapse is a locally naked singularity. We use the $1+1+2$ co-variant decomposition, explained above, to show this.\\\\
The Lagrangian of a scalar field is:
\be\label{Lag}
\mathcal{L}=-\frac12 g^{ab} \nabla_a \Psi \nabla_b \Psi - V(\Psi),
\ee 
where, $V(\Psi)$ is the potential associated to the scalar field and in the present case it is $zero$. Plugging this in the Hilbert-Einstein action and extremising it, we get the stress-energy tensor of a MSF:
\be \label{SE}
T_{ab}= \nabla_a \Psi  \nabla_b \Psi - -\frac12 g_{ab}  \nabla_c \Psi  \nabla^c \Psi. 
\ee
Euler-Lagrange's equation of motion or the Klein-Gordon equation, thus becomes:
\be \label{EL}
\nabla_a \nabla^a \Psi =0. 
\ee

\textcolor{black}{A MSF can have time-like, space-like and light-like gradients and it can be either of a type-I or a type-II matter field \cite{HE}. Choptuik's work, considers both the collapsing and dispersing modes and hence it is  convenient to formulate this problem in a double null-basis, consisting of radially incoming and outgoing null geodesics at each point \cite{Brady1995}.}\\

Null geodesics on $(\mathcal{M},g)$ are curves $x^{a}(\nu)$ whose tangents are parallelly propagated to themselves.
The tangent to these curves is defined by:
 \be\label{nullvector}
k^{b}\na_{b}k^{a} =\frac{dx^{a}(\nu)}{d\nu}= 0~,
\ee
where $\nu$ is an affine parameter along these geodesics. Now, a null-geodesic is outgoing if its tangent $k^a e_a>0$. Similarly, it is incoming if $l^a e_a<0$. Hence, the expression of outgoing null-geodesic along the preferred spatial direction (in this case, radially outgoing) can be written as:
\be\label{orng}
k^a = \frac{u^a + e^a}{\sqrt{2}}.
\ee
Using Eq.(\ref{del_e}) and Eq.(\ref{del_u}), it is easy to see that Eq.(\ref{nullvector}) is valid for the {\it outgoing radial null geodesic} (ORNG) (Eq.(\ref{orng})). {\it Incoming radial null geodesic} (IRNG) can similarly be constructed using:
\be\label{irng}
l^a l_a = l^b\nabla_b l^a =0\;\;\; \text{and}\;\;\; k^a l_a=-1. 
\ee
So, IRNG becomes:
\be
l^a = \frac{u^a - e^a}{\sqrt{2}}.
\ee 
The dynamics in the $2$-d {\it sheet} orthogonal to $k^a $ and $ l^a$, given by:
\be
\tilde{h}_{ab}= g_{ab} + 2 k_{(a}l_{b)}, \tilde{h}^a{}_a=2, \tilde{h}_{ab} k^a= \tilde{h}_{ab} l^a=0. 
\ee
Upon using the expressions of ORNG and IRNG in this equation, it is easy to see $\tilde{h}_{ab}=N_{ab}$. So, the formalism of $1+1+2$ splitting, explained above, is applicable if our objective is to follow the dynamics of the scalar field in the plane $[k,l]$. $k^a$ can now be split covariantly into its irreducible parts\cite{deSwardt:2010nf, Nzioki:2010nj, Ellis:2014, NF}:
\be\label{covk}
\nabla_bk_a=\sfr12 \tilde{h}_{ab}\tilde{\Theta}_{out}+\tilde{\sigma}_{ab}+\tilde{X}_ak_b+\tilde{Y}_bk_a
+\lambda k_ak_b,
\ee
where, $\tilde{X}_a=e^c\nabla_c k_a$, $\tilde{Y}_a=e^c\nabla_a k_c$ and $\lambda=-e^ae^b\nabla_a k_a$. $\tilde{\Theta}_{out}$ is the expansion scalar for the congruence of ORNGs and $\tilde{\sigma}_{ab}$ is the shear for the same. $\tilde{\Theta}_{out}$ can be calculated by contracting Eq.(\ref{covk}) with $\tilde{h}_{ab}$:
\be\label{thetao}
\tilde{\Theta}_{out}= \tilde{h}^{ab} \nabla_b k_a = \frac{N^{ab}}{\sqrt{2}} \nabla_a (e_b +u_b)
\ee
Using Eq.(\ref{del_e}) and Eq.(\ref{del_u}), we arrive at:
\be\label{Thetao}
\tilde{\Theta}_{out} = \frac{1}{\sqrt{2}} \left(\frac23 \Theta -\Sigma +\phi\right).
\ee
The same treatment, done with $l_a$, gives us the covariant derivative of it. In this case, the expansion of the congruence will be:
\be\label{Thetai}
\tilde{\Theta}_{in} = \frac{1}{\sqrt{2}} \left(\frac23 \Theta -\Sigma -\phi\right).
\ee\\
With these, we follow the evolution of the scalar field, whose gradient is decomposed in collapsing (along the IRNG congruence) and dispersing (along the ORNG congruence) modes:
\be\label{Psi}
\nabla_a \Psi = \Psi_+ k_a + \Psi_- l_a, 
\ee
where, $\Psi_-$ and $\Psi_+$ denote the strength of the collapsing and dispersing modes of the scalar field and they are functions of both $r$ and $t$.

It can be immediately seen that:
\be 
\nabla_a \Psi \nabla^a \Psi = -2 \Psi_+ \Psi_- .
\ee 
$\nabla_a \Psi$ is time-like if $\Psi_+$ and $\Psi_-$ have the same sign, $\nabla_a \Psi$ is space-like if they have the opposite signs. $\nabla_a \Psi$ is null if one of them vanish. 
$\nabla_a \Psi$ is given by:
\be \label{Psia}
\nabla_a \Psi= f u_a + g e_a\;\; ,
\ee
in the $[u,e]$ basis, with the assumption:
\be\label{fg}
f\equiv \frac{\Psi_+ +\Psi_-}{\sqrt{2}} \;\; , g\equiv \frac{\Psi_+ -\Psi_-}{\sqrt{2}}.
\ee
Using Eq.(\ref{Thermo_sc}), the thermodynamic scalars are:
\ba \label{coll}
\mu &=& \frac12 f^2 + \frac12 g^2\; , \nonumber\\
\label{collp}
p &=& \frac12 f^2 - \frac16 g^2\; , \nonumber\\
\label{collQ}
\mathcal{Q} &=& fg\; , \nonumber\\
\Pi &=& \frac23 g^2\;.
\ea
\textcolor{black}{We can easily see that the MSF obeys the equation of state: 
\be
\mu= p+\Pi\,.
\ee  
In the special case when the collapsing and dispersing modes exactly balance each other, that is $\Psi_+= \Psi_-$, then $\mu=p$ and $\Pi=\mathcal{Q}=0$ and the scalar field becomes a stiff fluid. }\\

\textcolor{black}{The energy momentum tensor (Eq.(\ref{STRESS})) can be written with space-time decomposition for the MSF (Eq.(\ref{SE})) in the following way:
\be\label{EM1}
T_{ab}=\mu u_au_b + (p+\Pi)e_ae_b +2\mathcal{Q}u_{(a}e_{b)}+(p-\frac12\Pi)N_{ab}\;.
\ee
To check the matter type of the MSF, we need to find the eigenvalues and eigenvectors of the above. Thus, by considering the eigenvalue equation\cite{Brassel:2021}: 
\be\label{eigen}
T_{ab} v^b =\lambda v_a,
\ee
where, $v_a$ is an arbitrary vector, we get the two equations corresponding to the $[u,e]$ plane from this:
\ba\label{eigenu}
T_{ab} u^a &=& -\mu u_b - \mathcal{Q} e_b\;,\\ 
\label{eigene}
T_{ab} e^a &=& (p+\Pi) e_b + \mathcal{Q} u_b\;. 
\ea
Using Eq.(\ref{coll}) in these equations, it is easy to see that only with $f=0$ or $g=0$, these are eigen-equations. For such a case, the energy-momentum tensor has a time-like eigen-vector with eigen-value $-\mu$, a space-like eigen-vector with eigen-value $(p+\Pi)$ and doubly degenerate space-like eigen-vectors with eigen-value $\left(p-\frac12 \Pi\right)$. For this case, the scalar-field is of type-I. $v^b$ can be replaced by $(u^a \pm e^a)$ in Eq.(\ref{eigen}) and from there it follows that:
\ba \label{eigenk}
T_{ab} (u^a + e^a) &=& (-\mu + \mathcal{Q}) u_b + (p+\Pi-\mathcal{Q}) e_b\;,\\
 \label{eigenl}
T_{ab} (u^a - e^a) &=& (-\mu - \mathcal{Q}) u_b - (p+\Pi+\mathcal{Q}) e_b\;.
\ea
Upon using Eq.(\ref{coll}) in these equations, we see iff $f=\pm g$, the energy-momentum tensor has double null eigen-vectors with eigen-value $(-\mu \pm \mathcal{Q})$. Therefore only in the special case of $f=\pm g$ (that is when either $\Psi_+=0$ or $\Psi_-=0$), the mass-less scalar field is of Type-II, otherwise in general it is a type-I field. As Choptuik considers both the collapsing and dispersing modes, therefore in this case the MSF is necessarily type-I.
}
\\\\
We now derive the Klein-Gordon equation for the scalar field by replacing Eq.(\ref{Psia}) in Eq.(\ref{EL}). Using the expressions of IRNG and ORNG in this, we get the Klein-Gordon equation in its irreducible form using Eq.(\ref{del_e}) and Eq.(\ref{del_u}):

\be\label{KGfg}
 \dot{f}+\hat{g}+f\Theta + g(\mathcal{A}+\phi)=0.
 \ee
We need the Ricci and the doubly contracted Bianchi identities to have the complete information about the evolution of the scalar field in the $[u,e]$ plane. With Eq.(\ref{del_e}) and Eq.(\ref{del_u}), we can derive the equations which specify directional derivatives along $\e_a$: \textit{propagation}, the derivatives along the world-lines of the fluid-flow (along $u_a$): \textit{evolution} and the equations which contain motion along both these directions: \textit{propagation/ evolution}. These are derived with doubly contracted Ricci identities and Bianchi identities and the detailed calculations are given in \cite{Clarkson:2002jz, Nzioki:2010, Clarkson:2007yp}. The following are these equations in terms of the scalar field dynamical quantities, given in Eq.(\ref{coll}):

\textit{Propagation}:
\ba \label{prop}
\hat\phi  &=&-\sfr12\phi^2+\bra{\sfr13\Theta+\Sigma}\bra{\sfr23\Theta-\Sigma}
    \nonumber\\&&-\frac12 f^2
    -\E ,\,\label{hatphinl}
\\  
\hat\Sigma-\sfr23\hat\Theta&=&-\sfr32\phi\Sigma-fg\
,\label{Sigthetahat}
 \\  
\hat\E-\sfr13(f\hat{f}-g\hat{g})&=&
    -\sfr32\phi\bra{\E+\sfr13 g^2}
    \nonumber\\&+&\bra{\sfr12\Sigma-\sfr13\Theta}fg.\;
\ea

\textit{Evolution}:
\ba
   \dot\phi &=& -\bra{\Sigma-\sfr23\Theta}\bra{\udot-\sfr12\phi}
+fg\ , \label{phidot}
\\   
\dot\Sigma-\sfr23\dot\Theta
&=&
-\udot\phi+2\bra{\sfr13\Theta-\sfr12\Sigma}^2\nonumber\\
        &&+\sfr13 (2f^2+g^2)-\E\, ,\label{Sigthetadot}
\\  
\dot\E -\sfr13(f\dot{f}-g\dot{g}) &=&
    \bra{\sfr32\Sigma-\Theta}\E
    +\sfr16\bra{\Sigma-\sfr23\Theta}g^2\nonumber\\
    &&-\sfr12\bra{f^2 +\frac{g^2}{3}}\bra{\Sigma-\sfr23\Theta} \nonumber\\
    &&+\sfr12\phi fg\ . \label{edot}
\ea

\textit{Propagation/evolution}:
\ba
   \hat\udot-\dot\Theta&=&-\bra{\udot+\phi}\udot+\sfr13\Theta^2
    +\sfr32\Sigma^2\nonumber\\ &&+f^2\ ,\label{Raychaudhuri}\\
\label{Bianchi1}
 f\dot{f} +g \dot{g}+ f\hat{g} + \hat{f} g &=& -\Theta (f^2 +\frac13 g^2) - (\phi + 2\mathcal{A}) f g\nonumber\\
 &&-\Sigma g^2 \;.
\\  \label{Bianchi2}
 f\dot{g} +g \dot{f} + f\hat{f} + g\hat{g} &=& - (\frac32 \phi +\mathcal{A})\frac23 g^2 -(\frac43 \Theta+ \Sigma) fg \nonumber\\
 &-& (f^2 +\frac13 g^2) \mathcal{A}
.\,
\ea
Taking (Eq.(\ref{Bianchi1})- $f\times$Eq.(\ref{KGfg})) and dividing it by $g$, we get:
\be\label{Bianchi}
\dot{g} + \hat{f} =-\left(\frac{\Theta}{3} +\Sigma\right) g -\mathcal{A} f ,
\ee
which we also get by taking (Eq.(\ref{Bianchi2})- $g\times$Eq.(\ref{KGfg})) and dividing it by $f$. Here, $f,g \neq 0$. 
Adding Eq.(\ref{KGfg}) and Eq.(\ref{Bianchi}); and subtracting Eq.(\ref{Bianchi}) from Eq.(\ref{KGfg}), we get the two governing equations that explain the dynamics of scalar field collapse:
\ba
\dot{\Psi}_+ +\hat{\Psi}_+ +\Psi_+ (\Theta + \mathcal{A}) &+& \nonumber \\
 + \frac{\Psi_+ -\Psi_-}{2} (\phi + \Sigma -\frac23 \Theta) &=&0\;,
 \label{psi+d} \\
\dot{\Psi}_- -\hat{\Psi}_- +\Psi_- (\Theta + \mathcal{A}) &+& \nonumber \\
 + \frac{\Psi_+ -\Psi_-}{2} (\phi - \Sigma +\frac23 \Theta) &=&0\; .  \label{psi-d} 
\ea 
With $\Psi_+, \Psi_- \neq 0$, these two equations only decouple on the apparent horizon and cosmological horizon \cite{NF}, given by:
\be\label{MOTS}
\Sigma - \frac23 \Theta = \mp \phi.
\ee
\textcolor{black}{So, we can never have $\Psi_+ =0$ or $\Psi_- = 0$ in this scenario, and therefore $f\neq \pm g$. Hence, once again we established that the MSF is strictly of Type-I \cite{HE}.} \\

\textcolor{black}{We would now like to clarify the following important point. If the MSF is strictly of Type I, then obviously there exists a special frame where the energy momentum tensor of the field is in the canonical diagonal form \cite{HE}. Hence, we should perform our collapse analysis in that special frame for better transparency, instead of working in a frame independent manner. However from Eq.(\ref{psi-d}), it is obvious that instead of one special frame, there are two special frames that brings the energy momentum tensor in the canonical diagonal form. The first one is the frame where $g=0$ ($\Psi_+=\Psi_-$) and the second one is the frame where $f=0$  ($\Psi_+=-\Psi_-$). It is obvious that both $f$ and $g$ cannot vanish simultaneously as in that case the energy momentum tensor will  identically vanish. It is important to note that each of these special frames restrict the causal nature of the field gradient in disconnected subsets. In other words the former frame restricts the gradient to be purely time-like, whereas the later frame restricts the gradient to be purely space-like. As Eq.(\ref{psi-d}) offers no such restrictions, hence none of these special frames are suitable to study the collapse of a MSF in a general setting. } 
 



\subsection{Collapsing scalar field and causal nature of  AH}\label{MisnerSharpAH}

Gravitational collapse starts from a regular initial space-like hypersurface which is called the initial configuration. The time-like congruence of a collapsing system, intersecting this 3D hypersurface, must be converging with $\Theta=D_au^a<0$. All the scalar, vector and tensor quantities on this surface must be well-behaved and integrable. If the system undergoes a UGC, then there can be a singular end-state with $\Theta \to -\infty$. \\
As discussed in the last section, the collapsing scalar field, here, is of type-I. Hence, there is a ''\textit{special}'' co-moving co-ordinate system, where its energy-momentum tensor is diagonal. But in the case of MSF, $\exists$ two such frames with null intersection: one where the gradient of the MSF is time-like $(g=0)$ and the other where the gradient of the MSF is space-like $(f=0)$. So, the gradient is restricted to the $[u,e]$ plane. For the case $g=0$, the time-like congruence with unit tangent $u^a$, is parallel to the gradient of the MSF. In the case $f=0$, the time-like congruence with unit tangent $u^a$, is orthogonal to the gradient of the MSF and the 2D sheet. So, the time-like congruence remains the same during the collapse with its $\Theta$ negative. To summarise, the choice of a timelike congruence is unique in a co-moving basis. In this basis, the MSF gradient is either along or perpendicular to the same in the [u,e] plane. In both of these cases, the end-state remains the same, which we will show later in this section.\\\\
We first check the regularity of the initial configuration. To do this, we employ Frobenius' integrability criterion\cite{wald}, which identifies an integrable submanifold $\subset (\mathcal{M},g)$. In spherical symmetry, the normal vectors $u^a$ and $e^a$ and their associated one-forms give the Frobenius' integrability criterion \cite{Nzioki:2010}:
\be \label{commutation}
\dot{\hat{\Upsilon}} - \hat{\dot{\Upsilon}} =\left(\frac{\Theta}{3}+\Sigma\right)\hat{\Upsilon}-\mathcal{A} \dot{\Upsilon}\;,
\ee 
where, $\Upsilon$ is a scalar defined on the plane $[u,e]$. This can also be seen as the Lie bracket of the vectors $\hat{\Upsilon}$ and $\dot{\Upsilon}$ lying in the same plane as the vectors themselves. This is another form of the Frobenius' integrability criterion.
In the current problem, these scalars are $f$ and $g$. These are integrable and well-defined on the $[u,e]$ plane and it can be seen by using Eq.s(\ref{KGfg},\ref{Bianchi},\ref{commutation}) and the propagation and evolution equations. On all the points $p\in \mathcal{M}$, these scalars will be well-behaved and integrable, despite our decomposition being local, because, spherical symmetry and time-orientabilty are global properties of the space-time. Thus, we find a regular initial configuration, defined by the quantities $f$ and $g$ or the collapsing and dispersing modes of the scalar field (Eq.(\ref{Psi})) or the thermodynamic quantities given in Eq.(\ref{coll}). This configuration, by virtue of the decomposition, is ansatz independent. So, there is no specific epoch where the collapse starts. Hence, all the thermodynamic scalars are well-behaved and integrable and finite all over $(\mathcal{V}, h_{ab})$ at the initial epoch of the collapse. During the course of the collapse,  {\it Apparent Horizon} (AH) or {\it marginally outer trapped surface} is formed. This is a 2-dimensional surface that acts as the boundary of the region containing all trapped surfaces in a 3-dimensional spatial slice \cite{Ashtekar:2002}. The causal restriction for the AH to be strictly space-like was removed 
in \cite{Ellis:2014} and this new definition is:\\
\begin{definition}
{\it A smooth, three-dimensional sub-manifold $\mathcal{H}$ in a spacetime $(\mathcal{M}, g)$
is said to be a Marginally outer trapped surface if it is foliated by a preferred family of $2$-spheres such that, on each leaf $\mathcal{S}$, the expansion $\tilde{\Theta}_{out}$ of the outgoing null
normal $k_a$ vanishes and the expansion $\tilde{\Theta}_{in}$  of the other ingoing null normal $l_a$ is
strictly negative} \cite{NF}.
\end{definition}\label{mots}
To locate the AH for a collapsing scalar-field, we now calculate the gravitational-mass enclosed in a shell of radius $\mathcal{R}$ at a specific time-instant. It was first calculated by Misner and Sharp \cite{Misner:1964} and called Misner-Sharp mass:
\ba\label{Misner}
\mathcal{M} = \frac{\mathcal{R}}{2} (1- \nabla_a \mathcal{R}\nabla^a \mathcal{R}),
\ea
where, $\mathcal{R}$ is related to the Gaussian curvature \hspace{2cm} $^2R_{ab}=KN_{ab}$, defined on the $2$-sheet as:
\ba
K &=& \frac{1}{\mathcal{R}^2} = \frac13 \mu - \mathcal{E} -\frac12 \Pi +\frac14 \phi^2 - {\left(\frac13 \Theta -\frac12 \Sigma\right)}^2,\nonumber\\ \label{GC1}
&=& \frac16 (f^2 -g^2)-\mathcal{E} +\frac14 \phi^2-\frac19 {\left(\Theta -\frac32 \Sigma\right)}^2,
\ea
where, Eq.(\ref{GC1}) is the Gaussian curvature of the collapsing scalar field. Like any other scalar, the covariant derivative of $\nabla_a K= -\dot{K} u_a + \hat{K} e_a$. The propagation and evolution equations are used to derive:

\ba\label{KK}
\dot{K} =\left( \Sigma -\frac23 \Theta\right) K ,\;\; \hat{K} =-\phi K.
\ea 
In our case, the MSF is undergoing a UGC and the Gaussian curvature must be increasing with time and $\dot{K}>0$ strictly. This implies $\Theta<0$, which is a necessary condition for collapse, as discussed before. Using Eq.(\ref{KK}) and Eq.(\ref{GC1}) in Eq.(\ref{Misner}), we get:
\ba\label{MS}
\mathcal{M}& =&\frac{1}{2\sqrt{K}} \left(1-\frac{1}{4K^3}\nabla_a K\nabla^a K\right), \\
\label{MS1}
&=& \frac{1}{2K^{\frac32}} \left(\frac16 (f^2-g^2) -\mathcal{E}\right).
\ea
For our current interest, we exclude shell-crossings, so $\forall \;\;\mathcal{R},\;\; \mathcal{\hat{R}}>0$. So, Eq.(\ref{KK}) implies $\hat{K}<0$ and $\phi>0$, as $K$ is a strictly non-negative quantity. So, in the $[u,e]$ plane, the AH curve is given by:
\be\label{AHd}
\varphi \equiv \frac23 \Theta + \phi -\Sigma =0.
\ee
With no shell-crossing, this definition ensures $\tilde{\Theta}_{in}<0$ and $\tilde{\Theta}_{out}=0$. We can now move on to calculate the Misner-Sharp mass function on the AH by calculating:
\ba\label{nablaK}
\nabla^a K \nabla_a K & =& -\dot{K}^2 + \hat{K}^2 \nonumber  \\
&=& \left(\frac23 \Theta -\Sigma + \phi\right) \left(\frac23 \Theta -\Sigma -\phi\right) \nonumber  \\
&=&0,
\ea 
which we get by using Eq.(\ref{KK}) and Eq.(\ref{AHd}). Using this and Eq.(\ref{MS}), we can identify the AH by:
\be\label{AH1}
\mathcal{M}=\frac{1}{2\sqrt{K}}\;.
\ee
With a clear definition of the AH (Eq.(\ref{AHd}) and Eq.(\ref{AH1})), we still need to understand the evolution of the same. This we do by assuming the tangent of $\varphi$ to be $\varphi^a = \alpha u^a +\beta e^a$ in the $[u,e]$ plane. We know that the gradient of the tangent to the AH will be parallel to itself, following Eq.(\ref{nullvector}), $\nabla^a \nabla_a \varphi =0$. With Eq.s (\ref{hatphinl}), (\ref{Sigthetahat}), (\ref{phidot}) and (\ref{Sigthetadot}), the slope is given by:
\ba
\frac\alpha\beta &=& -\frac{\hat{\varphi}}{\dot{\varphi}} \\
&=& \frac{\frac13 (f^2 + 2g^2)+\mathcal{E}-fg}{-\frac13 (2f^2 +g^2)+\mathcal{E}+fg}.\label{slope}
\ea
The AH is future outgoing if $\frac\alpha\beta >0$ and future incoming if $\frac\alpha\beta <0$. And a singularity is at least locally visible if the slope of the apparent horizon curve is $\geq 0$. In this case the AH is locally future outgoing.\\\\
Now, we calculate the slope of the AH for the collapsing scalar field, where we are considering the Misner-Sharp mass is enclosed within the AH. We take the expression of $\mathcal{E}$ from Eq.(\ref{MS1}) and replace it in Eq.(\ref{slope}) to get the slope of the AH in terms of the thermodynamic scalars:
\be\label{slopeAH}
Slope\equiv \mathcal{X}=\frac{\frac12 (f^2 + g^2) - 2\mathcal{M} K^{\frac32}-fg}{-\frac12 (f^2 + g^2)- 2\mathcal{M} K^{\frac32}+fg} 
\ee
Plugging back $\Psi_+$ and $\Psi_-$ in Eq.(\ref{slopeAH}), we get the slope of the AH for a MSF collapse:
\be\label{slopeML}
 \mathcal{X}=\frac{1-\frac{\Psi_-^2}{2\mathcal{M} K^{\frac32}}}{1+\frac{\Psi_-^2}{2\mathcal{M} K^{\frac32}}}\;.
\ee
To get rid of the Gaussian curvature, we replace the expression of Misner-Sharp mass on the AH from Eq.(\ref{AH1}) in this equation. It is to be noted that this expression, connecting Misner-Sharp mass to the AH, is a corollary of the Def.(\ref{mots}). So, $\mathcal{M}$ here denotes the geometric mass enclosed within the $2$-spheres, foliating the AH. Another important thing to note here is that Eq.(\ref{AHd}), which is written using Def.(\ref{mots}), decouples Eq.(\ref{psi-d}) from $\Psi_+$, which is the strength of the outgoing modes of the collapsing scalar field. So, essentially we see that, on the AH, the gradient of the scalar field is incoming null and the slope of AH is given by:
\be\label{AH2}
\mathcal{X}= \frac{1-{\eta}^2}{1+{\eta}^2},
\ee
where, $\eta = 2\mathcal{M} \Psi_-$. It is simple to see from this equation that $\eta$ is dimensionless. So, the slope and the evolution of the AH is independent of any length-scale. This shows the \textcolor{black}{scale-independent} nature of the MSF collapse, which Choptuik \cite{Choptuik} desscribes as "{\it echoing}". Also, this result is true for all frames and hence it is true for both of the co-moving frames where either $f$ or $g$ goes to \textit{zero}. This one parameter $\eta$, constructed by the strength of the collapsing modes of the scalar field and the Misner-Sharp mass on the AH, holds all the information about the visibility of the central singularity. So, the nature of $\eta$ is Universal, as predicted by Choptuik. Moreover, Choptuik's results hold true for all MSF collapse scenarios, even beyond the scalar-field families he considered. \\\\
Now, we follow the slope of the AH in the neighbourhood of the central singularity to comment on  the visibility of the same. \textcolor{black}{The collapse of the time-like congruence takes $\Theta \to -\infty$. We have already discussed that either $f=0$ or $g=0$ as the matter field is of type-I. Now, with Eq.(\ref{Bianchi1}) and $g=0$, it can be seen that $f= C\exp{[-\int\Theta dt]}$, where $C$ is the constant of integration. Hence, as $\Theta \to -\infty$, $f$ also blows up. With Eq.(\ref{coll}) it can be easily seen that, for this case, $\rho \to \infty$ and $p \to \infty$. The same occurs for $f=0$. In this case, $\rho \to \infty$, $p 
\to
-\infty$ and $\Pi \to \infty$. Thus, in both of these cases the physical quantities blow up and we encounter a genuine singularity in both of these cases. To analyse the visibility of this singularity, we now} move
{\it infinitesimally close} to the central singularity in the $[u,e]$ plane where, $t\to t_s\equiv$ time of formation of the central singularity, $r\to 0$ and the tangent of AH is $\mathcal{X}_0$ with $\eta\to \eta_0$, then: 
\be\label{X0}
\mathcal{X}_0= \frac{1-\eta_0^2}{1+\eta_0^2}\;,
\ee
where, $\eta_0 = 2\mathcal{M}_0 \Psi_{-0}$ and $\mathcal{M}\to \mathcal{M}_0$ is the Misner-Sharp mass and $\Psi_- \to \Psi_{-0}$ is the strength of collapsing modes of the scalar-field close to the singularity. So, we have three distinct outcomes for $\Psi_{-0} \in (0, \infty)$:\\
\begin{enumerate}
   \item 
{\bf Dispersal}: If $\Psi_{-0}$ is finite then none of  the dynamical quantities (Eq.(\ref{coll})) diverge as $r\to 0$ at any instant of time. So, there is no singularity formation.
\item 
{\bf Black-hole}: $\Psi_{-0}$ diverges fast then these quantities also diverge as $t\to t_s$ and $r\to 0$. So, there is a central singularity. Depending on the attained magnitude of $\eta_0$, we have two cases here: 
\begin{itemize}
\item {\bf a) $0<\eta_0 \leq 1 $}: In this case, $1>\mathcal{X}_0 \geq 0 $. So, the AH is outgoing space-like and the central singularity is covered. As we approach the singularity $r, \mathcal{R}\to 0$, Eq.(\ref{GC1}) tells that $K\to\infty$. So, the Misner-Sharp mass on the AH, $\mathcal{M}_0\to 0$, which we see from Eq.(\ref{AH1}). So, the central shell collapses at time $t_{s0}$ and gives rise to the critical $zero$-mass black-hole of Choptuik. As the collapse proceeds $(t>t_{s0})$, the mass of the central singularity starts growing up beyond zero.
\item {\bf b) $\eta_0 > 1 $}: Here, $\mathcal{X}_0 <0$ and the AH forms even before the formation of the singularity. In this case as well, $t_s$ is a function of the radial co-ordinate, as the scalar field has a non-zero gradient along the radial direction. And $t_{ah} < t_{s0}$, where $t_{ah}$ is the time of formation of the AH. So, the space-time diagram of the collapse is similar to Oppenheimer-Snyder-Datt dust (OSD) collapse. But the crucial difference between this case and OSD collapse is that, the whole cloud forms the central singularity at once in OSD collapse and in the present case, the collapsing configuration is inhomogeneous and the formation of the central singularity is not instantaneous.  
\end{itemize}
So, $\eta_0>0$,  always ends up in a black-hole.
\item
{\bf Locally naked singularity}: $\Psi_{-0}$ diverges slowly to take $\eta_{0}\to 0$. This is the critical boundary between cases {\bf (1)} and {\bf (2-a)}. $\mathcal{M}_0\to 0$ in this case as well. But in this case, as $\eta_{0}\to 0$, Eq.(\ref{X0}) gives $\mathcal{X}_0\to 1$. And we know that the slope to the AH in the neighbourhood of the singularity must be $\geq 1$ \cite{Goswami:2007, Koushiki:2025uqr}. In this case, we achieve the critical case of $\mathcal{X}_0\to 1$. In summary, we get a {\it zero-mass locally naked singularity} from the collapse of a MSF, whose gradient is along incoming null geodesic congruence at the AH. Therefore, this singularity is null in nature.
\end{enumerate}

In the cases where the end-state is singular, $\mathcal{M}_0 \to 0 \implies K\to \infty $ and Eq.(\ref{KK}) ensures $\Theta \to -\infty$. So, we self-consistently arrive at a system where the expansion of the time-like congruence is negative and it ultimately tends to negative \textit{infinity} at the end of the UGC.

\section{Discussion}

In this paper, we analytically reproduce critical collapse of a MSF in a frame-independent manner. We conclude that the strengths of collapsing and dispersing modes of the scalar field  will decide the final fate of the collapse, which might be either a regular or a singular state. We see that the behaviour of the AH, near the central singularity is dependent on a single dimensionless, hence, scale-invariant parameter $\eta$. This parameter is the product of the Misner-Sharp mass at the AH and the strength of the collapsing mode of the scalar field. Based on this, we have found out four different end-states to the collapse:\\
\textbf{a)} dispersal of the collapsing modes and a regular end state,\\
\textbf{b)} a black-hole, initially with zero gravitational mass,\\ 
\textbf{c)} a black-hole where the AH forms before the formation of the central singularity and\\
\textbf{d)} a locally naked null singularity with zero mass.\\
Thus, we have shown the existence of a critical  attractor in the phase space of $\eta$, for a general class of MSFs. By construction, our method is ansatz-independent and hence the measures of the subsets ending up in the black-hole and the naked singularity, are both non-zero.\\\\
 Although we have shown that this phenomenon is scale-independent, we have not studied the homothetic Killing symmetry near the threshold parameter value of the central singularity. This becomes important to show the attracting behaviour of the critical parameter in its phase-space. Physically, it is even more relevant because it shows that the length and time scales are irrelevant in such a collapse.  Brady \cite{Brady1995} has studied the self-similarity of critical scalar field collapse in double-null co-ordinates. But the initial choice of the co-ordinate chart $g_{ab}$, in his work entails self-similarity. So, probing the existence of homothetic Killing symmetry near the critical parameter still remains. We defer this work for future.


\begin{thebibliography}{99}


\bibitem{CCC}
R. Penrose, {\it Gravitational Collapse: The Role of General Relativity}, Riv. Nuovo Cimento, Num. Sp. I, 1969.

\bibitem{HE}
S. W. Hawking and G. F. R.
Ellis, \emph{\it The Large Scale Structure of Spacetime},
Cambridge University Press, 1973.


\bibitem{NF}
P.~S.~Joshi and D.~Malafarina, \emph{\it New Frontiers
in Gravitational Collapse
and Spacetime Singularities}, {\bf 255-271}
Springer, 2024.

\bibitem{wald} 
R.~M.~Wald, 
"{\it General Relativity}", 
University of Chicago Press, Chicago (1984).

\bibitem{ellis_van_elst}
G.~F.~R.~Ellis and H.~van~Elst,
"{\it Cosmological Models}",  Cosmology Group, Department of Mathematics and Applied Mathematics, University of Cape Town, Cape Town, South Africa (2008).

\bibitem{Choptuik}
M.~W.~Choptuik,
 {\it Universality and Scaling in Gravitational Collapse of a Massless Scalar Field},
 Phys. Rev. Lett. {\bf 70}, 1 (1993).

 \bibitem{Joshi:1991aa}
P.~S.~Joshi and I.~H.~Dwivedi,
``{\it Naked singularities in spherically symmetric inhomogeneous Tolman-Bondi dust cloud collapse},''
Phys. Rev. D \textbf{45}, 2147-2154 (1992).

\bibitem{Joshi:1993zg}
P.~S.~Joshi and I.~H.~Dwivedi,
``{\it The Structure of naked singularity in self-similar gravitational collapse},''
Commun. Math. Phys. \textbf{146}, 333-342 (1992).

\bibitem{Joshi:1994st}
P.~S.~Joshi and I.~H.~Dwivedi,
``{\it The Occurrence of naked singularities in spherically symmetric inhomogeneous perfect fluid collapse},''
Phys. Rev. D \textbf{47}, 5357-5369 (1993).

\bibitem{Joshi:1997kx}
P.~S.~Joshi and I.~H.~Dwivedi,
``{\it Naked singularities in Vaidya space-time},''
Phys. Rev. D \textbf{47}, 5357-5369 (1993).


\bibitem{Joshi:1993jg}
P.~S.~Joshi and R.~Maartens,
``{\it The Structure of naked singularities in selfsimilar gravitational collapse},''
Class. Quant. Grav. \textbf{10}, 1521-1532 (1993).

\bibitem{Goswami:2006ph} 
  R.~Goswami and P.~S. Joshi,
  `{\it Spherical gravitational collapse in N-dimensions},
  Phys.\ Rev.\ D {\bf 76}, 084026 (2007).
  [gr-qc/0608136].

\bibitem{Joshibook1} 
P.  S. Joshi, {\it Global Aspects in Gravitation and Cosmology}, Oxford University Press, 1993.

\bibitem{Joshibook2} 
 P. S. Joshi,{ \it Gravitational Collapse and Spacetime Singularities}, Cambridge University press, 2007.
 
\bibitem{Joshi:2001xi} 
  P.~S.~Joshi, N.~Dadhich and R.~Maartens,
  {\it Why do naked singularities form in gravitational collapse?},
  Phys.\ Rev.\ D {\bf 65}, 101501 (2002)
  [gr-qc/0109051].

\bibitem{Joshi:2004tb} 
  P.~S.~Joshi, R.~Goswami and N.~Dadhich,
  {\it Why do naked singularities form in gravitational collapse? 2}.,
  Phys.\ Rev.\ D {\bf 70}, 087502 (2004).

\bibitem{EllisLRS} G. F. R. Ellis: {\it The dynamics of pressure-free matter
in general relativity}. {\it Journ Math Phys} {\bf 8}, 1171 --
1194 (1967). H. van Elst  and G. F. R. Ellis, {\it Class. Quantum
Grav.} {\bf 13}, 1099 (1996), [gr-qc/9510044].

\bibitem{EllisCovariant} G.~F.~R.~Ellis \& H van Elst,
{\it Cosmological Models}, Carg\`{e}se Lectures 1998, in Theoretical
and Observational Cosmology, Ed. M Lachze-Rey, (Dordrecht: Kluwer
1999), 1. [arXiv:gr-qc/9812046].

\bibitem{Ellisbook} 
  George F. R. Ellis, Roy Maartens, Malcolm A. H. MacCallum 
 { \it Relativistic Cosmology} (Cambridge University press), 2007.

\bibitem{Clarkson:2002jz} 
  C.~A.~Clarkson and R.~K.~Barrett,
  {\it Covariant perturbations of Schwarzschild black holes},
  Class.\ Quant.\ Grav.\  {\bf 20}, 3855 (2003)
  [gr-qc/0209051].
  
\bibitem{Betschart:2004uu} 
  G.~Betschart and C.~A.~Clarkson,
  {\it Scalar and electromagnetic perturbations on LRS class II space-times},
  Class.\ Quant.\ Grav.\  {\bf 21}, 5587 (2004)
  [gr-qc/0404116].

\bibitem{Nzioki:2010}
A.~M.~Nzioki, S.~Carloni, R.~Goswami, and P.~K.~S.~Dunsby,
{\it New framework for studying spherically symmetric static solutions in $f(R))$ gravity},
Phys. Rev. D {\bf 81}, 084028 (2010).
  
\bibitem{Clarkson:2007yp} 
C.~Clarkson,
  {\it A Covariant approach for perturbations of rotationally symmetric spacetimes},
  Phys.\ Rev.\ D {\bf 76}, 104034 (2007).

\bibitem{Stewart-Ellis}
    J. M. Stewart J M and G. F. R. Ellis, 
    {\it On solutions of einstein's equations for a fluid which exhibit local rotational symmetry }\textit{J. Math. Phys.}
    {\bf 9} 1072, (1968)



\bibitem{deSwardt:2010nf} 
  B.~de Swardt, P.~K.~S.~Dunsby and C.~Clarkson,
  {\it Gravitational Lensing in Spherically Symmetric Spacetimes},
  arXiv:1002.2041 [gr-qc].
  
 
 
  
  
\bibitem{Nzioki:2010nj} 
  A.~M.~Nzioki, P.~K.~S.~Dunsby, R.~Goswami and S.~Carloni,
  ``A Geometrical Approach to Strong Gravitational Lensing in f(R) Gravity,''
  Phys.\ Rev.\ D {\bf 83}, 024030 (2011)
  [arXiv:1002.2056 [gr-qc]].
  

\bibitem{GosEllis} 
  R.~Goswami and G. F. R. Ellis,
  ``{\it Almost Birkhoff Theorem in General Relativity}'',
  Gen.\ Rel.\ Grav.\  {\bf 43}, 2157 (2011)
  [arXiv:1101.4520 [gr-qc]].

\bibitem{senovilla} 
J. M. M. Senovilla, ``{\em Remarks on the Stability Operator for MOTS}", Progress in Mathematical Relativity, Gravitation and Cosmology,
Springer Proceedings in Mathematics and Statistics Volume 60, 403 (2014).


\bibitem{psj} 
  P.~S.~Joshi,
  ``{\em Visibility of a spacetime singularity}'',
  Phys.\ Rev.\ D {\bf 75}, 044005 (2007).
  
  
\bibitem{Goswami:2004gy} 
  R.~Goswami and P.~S.~Joshi,
  ``{\em Cosmic censorship in higher dimensions}'',
  Phys.\ Rev.\ D {\bf 69}, 104002 (2004)
  [gr-qc/0405049].

\bibitem{RG}
 R.~Goswami and P.~S.~Joshi,
{\it Spherical gravitational collapse in N-dimensions},
 Phys.\ Rev.\ D {\bf 76}, 084026 (2007).

\bibitem{Goswami:2002he} 
  R.~Goswami and P.~S.~Joshi,
  ``{\em Spherical dust collapse in higher dimensions}'',
  Phys.\ Rev.\ D {\bf 69}, 044002 (2004)
  [gr-qc/0212097].



 
  
\bibitem{Zibin:2008vj} 
  J.~P.~Zibin,
  ``{\em Scalar Perturbations on Lemaitre-Tolman-Bondi Spacetimes}'',
  Phys.\ Rev.\ D {\bf 78}, 043504 (2008)
  [arXiv:0804.1787 [astro-ph]].

\bibitem{Kong:2013daa} 
  L.~Kong, D.~Malafarina and C.~Bambi,
  ``{\em Can we observationally test the weak cosmic censorship conjecture?}'',
  arXiv:1310.8376 [gr-qc].

\bibitem{Misner:1964}
C.~W.~Misner, D.~H.~Sharp, 
``{\em Relativistic Equations for Adiabatic, Spherically Symmetric Gravitational Collapse}''
Phys.\ Rev.\ D {\bf 136}, B571 (1964).

\bibitem{Ellis:2014}
G.~F.~R.~Ellis, R.~Goswami, A.~I.~M.~Hamid, S.~D.~Maharaj, 
``{\em Astrophysical black hole horizons in a cosmological context: Nature and possible consequences on Hawking radiation}''
Phys.\ Rev.\ D {\bf 90(8)}, 084013 (2014).

\bibitem{Hamid:2017}
 A.~I.~M.~Hamid, R.~Goswami and S.~D.~Maharaj,
``{\em Cosmic Censorship Conjecture revisited: Covariantly}''
The Fourteenth Marcel Grossmann Meeting. December 2017, 2615-2620.

\bibitem{Koushiki:2025uqr}
Koushiki and P.~S.~Joshi,
``{\em Apparent horizon and causal structure of spacetime singularities}''
[arXiv:2508.14663 [gr-qc]].


\bibitem{Ashtekar:2002}
A.~Ashtekar, B.~Krishnan, 
``{\em Dynamical Horizons: Energy, Angular Momentum, Fluxes and Balance Laws}''
Phys. Rev. Lett. 89, 261101 (2002).

\bibitem{Christ}
D.~Christodoulou, 
"{\it The formation of black holes and singularities in spherically symmetric gravitational collapse}"
Commun.Pure Appl.Math. {\bf 44} 3, 339-373 (1991).

\bibitem{Christ1}
 D.~Christodoulou,
 "{\it The Instability of Naked Singularities in the Gravitational Collapse of a Scalar Field}"
 Annals of Mathematics, {\bf 149}, 183
(1999).

\bibitem{Goswami:2007}
R.~Goswami, P.~S.~Joshi,
"{\it Gravitational collapse of a self-interacting scalar field}"
Mod.Phys.Lett.A {\bf 22}, 65-74 (2007).

\bibitem{KK1}
K.~Mosani, K., P.~S.~Joshi, J.V.~Trivedi and T.~Bhanja,
"{\it Gravitational collapse of scalar and vector fields}"
Phys. Rev. D \textbf{ 108}, no.4,  044049 (2023).

\bibitem{Brassel:2021}
B.~P.~Brassel, S.~D.~Maharaj and R.~Goswami,
"{\it Inhomogeneous and Radiating Composite Fluids}"
 Entropy  \textbf{23}, no.11, 1400 (2021).

\bibitem{DeyK:2023}
D.~Dey, K.~, and P.~S.~Joshi,
"{\it Equilibrium states from gravitational collapse of a minimally coupled scalar field with a nonzero potential}" Phys. Rev. D \textbf{108}, 104045 (2023).

\bibitem{Tipler} F. J. Tipler, {\it Singularities in conformally flat spacetimes} Phys. Lett. \textbf{64}A, 8 (1977).

\bibitem{JMN11} 
P. S. Joshi, D. Malafarina, and R. Narayan, 
``{\it Equilibrium configurations from gravitational collapse},''
Class. Quantum Grav. {\bf 28}, 235018 (2011).

\bibitem{Joshi:2013dva}
P.~S.~Joshi, D.~Malafarina and R.~Narayan,
``{\it Distinguishing black holes from naked singularities through their accretion disc properties},''
Class. Quant. Grav. \textbf{31}, 015002 (2014).

\bibitem{Abrahams}
A.~M.~Abrahams and C.~R.~Evans
"{\it Critical behavior and scaling in vacuum axisymmetric gravitational collapse},"
Phys. Rev. Lett. \textbf{70}, 2980 (1993).

\bibitem{Hamad:2004sw}
R.~S.~Hamade and J.~M.~Stewart,
"{\it Critical Phenomena in Gravitational Collapse},"
Class. Quant. Grav. \textbf{13}, 497-512 (1996).

\bibitem{Garfinkle:1995zj}
D.~Garfinkle and G.~C.~Duncan, "{\it Scaling of curvature in subcritical gravitational collapse},"
Phys. Rev. D, \textbf{58}, 064024 (1998).

\bibitem{Hod:1996ar}
"S.~Hod and T.~Piran,
{\it Fine structure of Choptuik's mass scaling relation}," Phys. Rev. D, \textbf{55}, R440(R) (1997).

\bibitem{Garfinkle1998}
D.~Garfinkle and G.~C.~Duncan,
{\it Scaling of curvature in subcritical gravitational collapse},"
Phys. Rev. D, \textbf{58}, 064024 (1998).

\bibitem{Gundlach:2002sx},
C.~Gundlach, "{\it Critical phenomena in gravitational collapse: Living Review},"
Living Rev. Rel., \textbf{7}, 5 (2004).

\bibitem{Brady1995}
P.~R.~Brady,
"{\it Analytic example of critical behavior in scalar field collapse},"
Class. Quant. Grav. \textbf{11}, 1255-1260 (1995).

\bibitem{Carr2021}
B.~Carr, F.~Kuhnel,
"{\it Primordial black-holes as dark matter: Recent developments},"
Annu. Rev. Nucl. Part. Sci. \textbf{70}: 355-94 (2020).

   
\end{thebibliography}
\end{document}